\documentclass[conference, 10pt]{IEEEtran}
\IEEEoverridecommandlockouts

\usepackage{cite}
\usepackage{amsmath,amssymb,amsfonts}
\usepackage{graphicx}
\usepackage{textcomp}
\usepackage{xcolor}
\def\BibTeX{{\rm B\kern-.05em{\sc i\kern-.025em b}\kern-.08em
    T\kern-.1667em\lower.7ex\hbox{E}\kern-.125emX}}

\usepackage{graphicx,latexsym,amsfonts,amsmath,amsthm,amssymb,verbatim,multicol,babel,subfigure}

\newtheorem{lemma}{Lemma}
\newtheorem{definition}{Definition}
\newtheorem{example}{Example}
\newtheorem{prof}{Proof}
\usepackage{empheq}
\usepackage[normalem]{ulem}
\usepackage{algorithm}
\usepackage{algpseudocode}
\usepackage{amsmath, amssymb}

\begin{document}

\title{Optimizing Resource Allocation in a Distributed Quantum Computing  Cloud: A Game-Theoretic Approach*\\

}

\author{\IEEEauthorblockN{1\textsuperscript{st} Bernard Ousmane Sane}
\IEEEauthorblockA{\textit{Graduate School of Media } \\
\textit{and Governance, Keio University}\\
Fujisawa, 252-0882, Japan \\
bernard@sfc.wide.ad.jp
}
\and
\IEEEauthorblockN{2\textsuperscript{nd}Michal Hajdušek }
\IEEEauthorblockA{\textit{Graduate School of Media} \\
\textit{and Governance, Keio University}\\
Fujisawa, 252-0882, Japan \\
michal@sfc.wide.ad.jp}
\and
\IEEEauthorblockN{3\textsuperscript{rd} Rodney Van Meter}
\IEEEauthorblockA{\textit{Graduate School of Media} \\
\textit{and Governance, Keio University}\\
Fujisawa, 252-0882, Japan \\
rdv@sfc.wide.ad.jp}

}
\maketitle

\begin{abstract}
Quantum cloud computing will play a vital role in achieving quantum supremacy by providing flexible allocation of resources for applications that demand extensive computational power. Such a platform allows various clients to run their quantum jobs (quantum circuits) without managing the quantum hardware and to pay based on resource usage. Therefore, defining optimal quantum resource allocation is essential to prevent overcharging clients and to help quantum cloud providers maximize resource utilization. Our approach involves examining the problem from a game theory perspective. We introduce a quantum circuit partitioning resource allocation game model (QC-PRAGM) designed to minimize client costs and inter-node communication. This is achieved by solving a convex optimization problem that finds the system’s optimal cost while maximizing the number of local gates within a partition through the selection of the best qubit combinations. We demonstrate analytically that clients are charged fairly. Additionally, our simulations show that our solutions outperform traditional methods.
\end{abstract}


\section{Introduction}
Cloud computing, born from the evolution of information systems, has transformed computing by enabling companies and individuals to access computing services at a lower cost through providers. The same approach extends to quantum computing, with providers (IBM Quantum, IonQ, Amazon Braket, Rigetti, and Azure Quantum) offering accessible and affordable quantum computing environments \cite{Gonzalez2021ER}. Today's quantum computers are highly specialized, delicate machines. They are hence installed and managed primarily in dedicated facilities and accessed via a web-based or Internet protocol-based front end, with services provided to users all around the world.  Thus, these systems are often called \emph{cloud-based quantum systems}, or are said to provide \emph{quantum computing as a service} (QCaaS)~\cite{grossi2021serverlesscloudintegrationquantum}. Consequently, customers do not need their own quantum machine and a suitable environment to execute their quantum algorithms. Instead, users can take advantage of vast computing power with just their classical machines and Internet access. They can create and submit jobs through a front-end management system that handles queue management, resource allocation, compilation, execution, and data collection. (Data pre- and post-processing may be done at either the user's premises or at the quantum provider's premises; for our purposes in this paper, the difference is not important.) This is crucial to democratizing quantum computing. 

The resource requirements for post-classical quantum computation, particularly the number of qubits needed, can vary significantly. For some quantum chemistry and cryptographically relevant problems, the required number of physical qubits can escalate to millions \cite{Gidney2021howtofactorbit,van-meter04:fast-modexp}. Today's NISQ-era systems range in capacity from only a few physical qubits to nearly a thousand. So it has been proposed that these systems be multi-programmed to execute multiple independent jobs at the same time, using independent subsets of the systems' qubits~\cite{das19:multiprog}. To date, the execution of a single quantum circuit has been confined to a single quantum computer, like in NISQ-era systems. There are, however, several hardware constraints that prevent the implementation of more than a few physical qubits on each QPU. Therefore, distributed computing is fundamentally important in competing with large-scale computing tasks in quantum cloud environments. It enables large and complex quantum tasks to be divided into manageable parts, which are then distributed among interconnected quantum nodes (QPUs). This is important for processing complex algorithms efficiently as well as promoting scalability. Over the last twenty years, researchers have proposed coupling smaller quantum computers using an entangling network to create larger systems known as \emph{multicomputers}, capable of solving problems exceeding the capacity of individual nodes~\cite{van-meter06:thesis}.  Recently, the first demonstrations of remote gate execution have been shown~\cite{main2025distributed}.
Various commercial quantum computing companies have near-term road maps toward fault tolerance that end around 2030, with the largest single system they can build (which we call ``MaxSQ'').  Thus, in the early part of the next decade, we anticipate that multicomputer approaches will become not only common but necessary.

Recent network design, Q-Fly~\cite{sakuma2024opticalinterconnectmodularquantum}, suggests that, like modern supercomputers \cite{kim2009dragonfly}, it will be possible to scale data center networks of quantum nodes to tens or even hundreds of thousands of nodes using group-based approaches building on Dragonfly networks. The limiting factor in such a design will be the per-switch photon loss.

With limited scaling of multicomputer or data center-sized systems available in the foreseeable future, it becomes important to utilize those resources effectively.  Some jobs can be expected to utilize the entire system, but others will need only a small fraction of the total nodes available.  While some jobs will fit within individual nodes, many others will require resources spanning subsets of the system.  Thus, like parallel job execution in large supercomputers, it becomes important to efficiently allocate resources to a set of jobs, including sometimes sharing parts of individual nodes among two or more jobs~\cite{downey97:par-workload}.  It is this problem that we address in this paper.

\begin{figure*}
    \centerline{\includegraphics[scale=0.35]{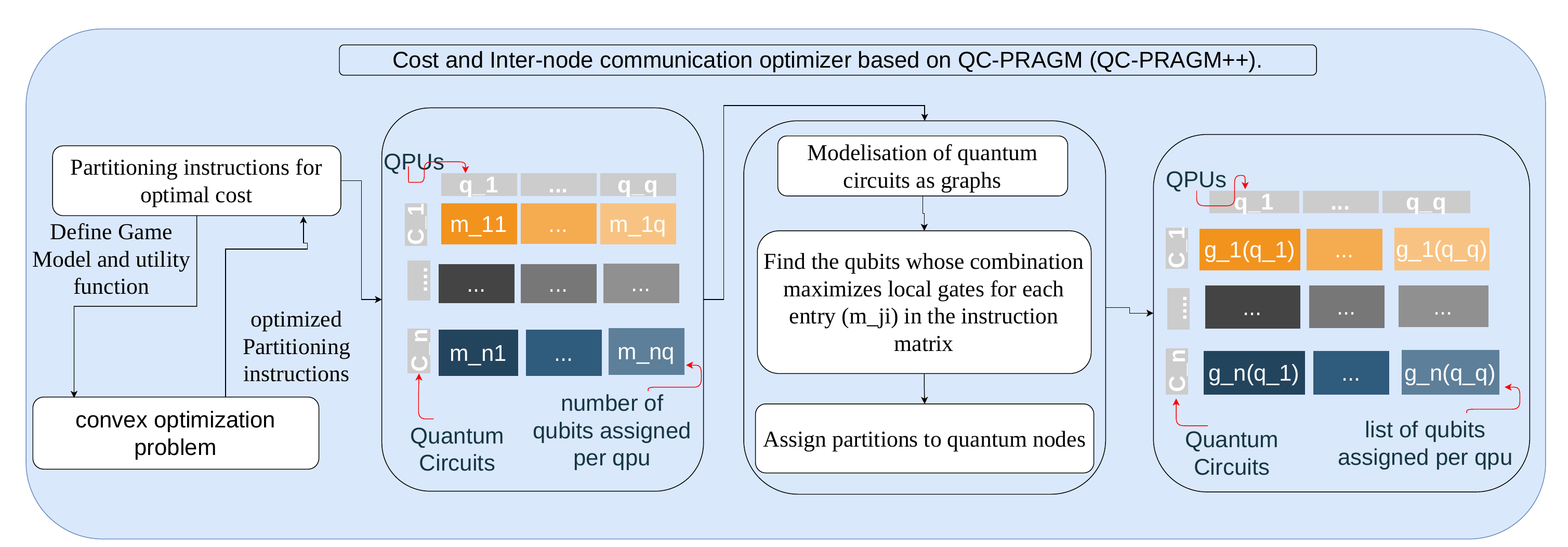}}

    \caption{A diagram illustrating our solution. Using the game model, we derive a cost-optimized partitioning instruction represented by a matrix ($M$) where each row ${m_{j1}, \dots, m_{jq}}$ represents the resources (number of qubits) of circuit $j$ to allocate on $q$ quantum nodes. However, this matrix only provides a numerical distribution of qubits. We, therefore, add a process to optimize their grouping, which minimizes costs and improves inter-node communication efficiency. This process is based on a graph representation of circuits, introducing a metric based on the total weight of vertex pairs (qubits). For each $m_{jk}$ in the matrix ($M$), we search for a list of qubits of size $m_{jk}$ maximizing this weight, i.e., containing the most local gates. Finally, we obtain a matrix ($\mathcal{G}$) in which each row ${g_j(q_1), \dots, g_j(q_q)}$ represents the resources (list of qubits) to allocate from circuit $j$ across $q$ quantum nodes, thus defining the client's strategy. Therefore, ($\mathcal{G}$) outlines the partitioning instructions that minimize costs and improve communication, allowing quantum nodes to be assigned specific partitions.}
     
    \label{fig:Overview-of-Our-Proposal}
\end{figure*}

\subsection*{Motivations, novelty and quantum suitability}
In a quantum cloud computing environment, we have multiple clients with different resource requirements, and the provider is expected to meet the clients’ needs by allocating appropriate quantum computers to them while optimizing the available resources. Moreover, clients are supposed to be billed only for the resources they consume (e.g., based on QPU hours), but existing allocation techniques do not support this factor. Consequently, the client can be unfairly charged. To the \textit{best of our knowledge, there is no existing work that addresses client costs in quantum job execution in a multi-user cloud environment}. 

Therefore, we explore a classical cloud resource allocation method for pricing optimization \cite{articleJalaparti}, which we adapted to the specific needs of the quantum cloud to manage not only pricing in the quantum cloud but also the management of quantum circuit partitions in distributed systems. It should also be noted that the allocation of resources for quantum cloud computing presents much more complex challenges than for classical computing. There are several reasons for this, including the sensitivity of qubits, the imperfection of quantum hardware, connectivity, etc. Our approach is based on game theory, as demonstrated by the studies in \cite{San2024InterdependencyAS}, which show that game theory is effective in traditional cloud environments for addressing resource allocation problems. 

\subsection*{Contributions}
We apply classical game theory to the problem of resource (circuit) allocation to optimize quantum job execution in a multi-user cloud environment. Hence,
\begin{itemize}
    \item
    We propose the first resource allocation game model that addresses client pricing and inter-node communication in the quantum cloud. 
    \item Our algorithm simultaneously addresses inter-node communication, reduces costs, and optimizes resource usage at the cloud level.
    \item We determine the number of partitions for each quantum circuit submitted by the clients and allocate them to the appropriate quantum nodes.
         
\end{itemize}

Analytically, we show that clients are charged appropriately (at a cost no higher than $\frac{4}{3}$ the optimal cost) while optimizing quantum cloud resources. We assess our proposal by considering a completely connected network and choosing quantum circuits from three distinct types: Quantum Fourier Transform (QFT), Deutsch-Jozsa (DJ), and GHZ state preparation. The simulation results demonstrate that our proposal is efficient for circuit partitioning in a distributed environment. It outperforms the round robin and random algorithms regarding the cost per quantum node, total cost, maximum cost, number of partitions, number of remote gates, and in handling latency-related
errors. The overview of our solution is presented in Figure~\ref{fig:Overview-of-Our-Proposal}.

\section{System model and problem definition}
\label{sec:system-model-and-problem-def}

 
We consider Quantum Cloud Computing (QCC) with multiple quantum nodes interconnected via both quantum and classical links.
The QCC provides computational resources on a pay-per-use basis.
Users can submit a variety of quantum circuits, each with different requirements, known as jobs.
Based on the resource requirements of each job, the quantum cloud provider makes allocation decisions.
Following these decisions, the quantum circuits are partitioned and assigned to different quantum nodes.
Complex jobs can be divided into smaller, more manageable parts, which are then processed across the quantum nodes.

 
 
 \subsection*{Mathematical model}
      
The quantum cloud provider offers a platform that includes a collection of quantum computers to meet the diverse needs of its customers in quantum computing. This collection, denoted as $ Q = \{ q_1, \ldots, q_q \} $, consists of $ q $ quantum computers, which are represented as quantum nodes. These nodes are available during the allocation period and collectively form a QCC system, providing QPU resources. QPUs are the main resource at quantum nodes, and they are described by their capacity (number of computing qubits). Quantum node and QPU are interchangeable throughout the paper. 


The batch of clients' quantum circuits is given by a set $\mathcal{C} = \{C_1, \ldots, C_n\}$ where $C_j$ represents a job (quantum circuit) from client $j$. Each $C_j$ requests $r_j$ resources from quantum cloud computing. Where $r_j$ highlights the number of computing qubits required to execute the circuit $C_j$.  The total client requests are represented by $\mathcal{R} = (r_1, \ldots, r_n)$ (resource requirement vector). 

The quantum cloud computing allocates a global resource $\mathcal{G} = (g_1, \ldots, g_n )$ to every client, where $g_j$ represents the total resources allocated to a client $j$ over the $q$ quantum nodes, in this sense representing the client's strategy in the game. In this case, $g_j$ is equal to $(g_j(q_1), \dots, g_j(q_q))$, where $g_j(q_k)$ indicates the resource allocated to the client $j$ at quantum node $q_k$. In other words, $g_j = (g_j(q_1), \dots, g_j(q_q))$ indicates how user $j$'s quantum circuit is partitioned among the available quantum nodes. Therefore, for $q$ quantum nodes and $n$ quantum circuits, a possible allocation can be illustrated by the following matrix $\mathcal{G}$ 
\begin{equation}
 \begin{array}{c|cccc}
      &q_1 & q_2 & \ldots & q_q \\ \hline
   C_1 &g_1(q_1) & g_1(q_2) & \ldots & g_1(q_q) \\
   \vdots & \vdots & \vdots & \vdots & \vdots \\
   C_n &g_n(q_1) & g_n(q_2) & \ldots & g_n(q_q)
\end{array}
\end{equation}

\subsection{Game model}
Game theory is a mathematical tool that models interactions between participants, known as players, to ensure the optimal payoff for each. There are various types of games, but in all of them, we find three concepts: players, their strategies, and the payoff associated with these strategies. The primary goal of each player in the game is to choose the best strategy to maximize their gains. Therefore, we aim to reach an equilibrium state known as a Nash equilibrium (Def. \ref{def:nash}).

In our quantum circuit partitioning resource allocation game model (QC-PRAGM), the interactions between quantum circuits of clients sharing the same quantum resource and inter-node communication are efficiently managed, ensuring optimal utilization of resources through this model. Based on Tim Roughgarden et \textit{al}. paper \cite{10.1145/506147.506153}, we prove that the Nash equilibrium (Def. \ref{def:nash}) exists. This model treats clients as players in the game, and resource allocation represents their strategies. We consider the following assumptions.
\begin{itemize}
   
    \item Circuits are aligned by priority order, considering factors such as deadlines, dependencies on other tasks, etc.

     \item Once allocated to a quantum node, each task executes until its completion, without interruption.
    \item Quantum nodes can host only one type of resource (for example, QPU), and clients are only concerned about using this resource type.
    \item The clients are selfish and are charged based on their QPU consumption. On the other hand, the provider of the quantum cloud aims to optimize resource usage.
    \item  The quantum processing units (QPUs) differ from each other based on their capacities.
\end{itemize}

A quantum circuit can take longer to execute in a heavily loaded quantum cloud computing environment due to the interactions between client quantum circuits sharing the same quantum resources. As a result, the client can be unfairly charged. Hence, each quantum node $q_k \in Q$ is associated with a cost function $f(q_k, x)$ defined as follows
\begin{equation}
    \begin{split}
       f: \mathbb{R}^{+} \rightarrow  Q\times\mathbb{R}^{+}. \\
       x \rightarrow f(q_k, x)
    \end{split}
    \label{equ:cost}
\end{equation}
 It assigns the clients' cost at quantum node ${q_k}$ (which has $m_k$ available resources) using $x$ resources. The function $f(q_k, x)$ represents the amount of money relative to the number of computing qubits used by the clients' circuit at quantum node $q_k$. It is proportional to the completion time (QPU-hours) when consuming $x$ resources at the quantum node $q_k$. In practice, it corresponds to the amount of time it takes for the clients' circuits to use computing qubits at QPUs. 
 \begin{equation}
    \begin{split}
       W({V_{g_j(q_k)}}) = \sum_{\{v_i, v_j\} \in V_{g_j(q_k)}} w(v_i, v_j)
    \end{split}
    \label{equ:weight}
\end{equation}
 In addition to the cost function \( f \), it is also important to consider the communication between the nodes. To address this, we define a metric aimed at reducing the number of remote gates by maximizing the number of local gates within each group of qubits, denoted as \( g_j(q_k) \). Indeed, one challenge in implementing distributed quantum circuits is the use of non-local gates (remote gates), which necessitate communication between qubits situated on different QPUs. In this part, we focus on identifying the optimal qubit combinations of size \( |g_j(q_k)| \)  that maximize the number of local gates. This optimization does not change the total number of qubits assigned to each node. We know that the number of remote gates remains unknown until the partition becomes effective, which occurs after resource allocation. To address this issue and minimize the number of remote gates, we represent the circuit $C_j \in \mathcal{C}$ as a graph $(V_j, E_j)$. In this graph, vertex $V_j$ represents qubits, while edges $E_j$ represent two-qubit gates. We employ $w(v_i, v_j)$ to map a pair of vertices to their corresponding weights (the number of gates) associated with the given edge. We aim to maximize the total weight of vertex (qubit) pairs in $g_j(q_k)$. Assuming the set of vertex pairs in \( g_j(q_k) \), denoted as \( V_{g_j(q_k)} \), we formulate the total sum of the weights associated with \( V_{g_j(q_k)} \) in \eqref{equ:weight}.

During the allocation period, the quantum cloud provider receives a set \( Q = \{ q_1, \ldots, q_q \} \), which consists of \( q \) quantum nodes, each defined by the number of computing qubits it possesses. Additionally, the provider receives all client requests \( \mathcal{R} = (r_1, \ldots, r_n) \), which represent the resource requirement vectors. With these inputs, there are various ways to allocate resources without exceeding the available resources at the quantum nodes. In our resource allocation approach, we aim to minimize client costs and ensure optimal inter-node communication for better resource utilization, thereby defining utility functions. 


   

\begin{definition} (Client's utility) Let $
\mathcal{G}$ be the resource allocation,
$j$ designates the player (client); the client's objective is to minimize the cost \eqref{equ:client-cost} and maximize the number of local gates; their utility is
\begin{equation}
    \begin{split}
       \mathcal{U}(\mathcal{G},j) = -\alpha\mathcal{F}(\mathcal{G},j) + (1 - \alpha)\mathcal{T}(\mathcal{G},j) \\
    \end{split}
    \label{equ:client-utility}
    \end{equation}
Where:
\begin{equation}
    \begin{split}
       \mathcal{F}(\mathcal{G},j) = \sum^{q}_{k=1} \mathcal{E}(j, q_k) f(q_k, \sum^{n}_{j=1}g_j(q_k)) \\
    \end{split}
    \label{equ:client-cost}
\end{equation}
\begin{equation}
    \begin{split}
    \mathcal{T}(\mathcal{G},j) = \sum^{q}_{k=1} \mathcal{E}(j, q_k)\sum^{n}_{j=1}
       W_{V_{g_j(q_k)}}
    \end{split}
    \label{equ:weight2}
\end{equation}

Here, $\mathcal{E}(j, q_k)$ is an indicator equal to $1$ if quantum node $q_k$ is used to allocate user $j$ circuits, $f(q_k, \sum^{n}_{j=1}g_j(q_k))$ designates the clients' cost at quantum node ${q_k}$, $\sum^{n}_{j=1}
       W(V_{g_j(q_k)})$ represents the number of local gates in the quantum node $q_k$, and $\alpha$ represents the weight associated with clients' costs and inter-node communication based on their importance.

\end{definition}


In the context of $\mathcal{G}$, the performance of the quantum cloud is assessed by the total cost incurred in the system. The provider's goal is to maximize the utility associated with this system \eqref{equ:system-cost}. 
\begin{equation}
    \begin{split}
        \mathcal{S}(\mathcal{G}) = -\alpha\sum^n_{j = 1}  \mathcal{F}(\mathcal{G},j) + (1 - \alpha)\sum^n_{j = 1}\mathcal{T}(\mathcal{G},j)
    \end{split}
    \label{equ:system-cost}
\end{equation}


\begin{definition}
The quantum circuit partitioning resource allocation game model (QC-PRAGM) is defined by $(\mathcal{Q}, \mathcal{U}, \mathcal{R})$. The goal of QC-PRAGM is to allocate clients' quantum circuits in quantum cloud computing efficiently. Each client, denoted as client \( j \), aims to minimize the costs associated with job interactions and inter-node communications. This is achieved by maximizing their utility, as described by equation \eqref{equ:client-utility}. In contrast, the quantum cloud provider seeks to maximize the system's performance by maximizing \eqref{equ:system-cost}.  In addition, QC-PRAGM models should meet the following constraints:
\begin{itemize}
  
    \item Minimizing the number of partitions by promoting more zeros in $g_j$. In other words, the resource $g_j = (g_j(q_1), \dots, g_j(q_q))$ allocated to client $j$ should be sparse with respect to the $l_1-$norm $|| x || = \sum^{p}_{i = 1}x_i$ \cite{bach2011convex} (constraint $C_1$).
    \item The total resources allocated to a client across all $q$ quantum nodes equals their demand. Specifically, for all $j$, $\sum^{q}_{k=1}g_j(q_k) = r_j$ (constraint $C_2$).
    \item The total resources the clients' circuits use in a single quantum node are less than those available at that node. Specifically, for all $k$, $\sum^{n}_{j=1}g_j(q_k) \leq m_k$ (constraint $C_3$).
\end{itemize}

\label{def:QC-PRAGM}
\end{definition}

Our game model assumes that clients act in their self-interest; they aim to maximize their utility as described in equation \eqref{equ:client-utility}. Consequently, each client $j$ will adjust its strategy $g_j$ as much as possible to achieve better utility. However, this may negatively impact other clients. This battle will continue until the system reaches the Nash equilibrium. It is the state of the game in which no client can unilaterally alter its strategy and achieve a better outcome. As a formal definition, we provide the following:

\begin{definition}[Nash equilibrium] 
Let $G’_j$ represent the collection of all possible strategies for the client $j$, where $j=1, \ldots, n$, the allocation of resources $(g_1, \ldots, g_n )$ is a Nash equilibrium if 

\[
\mathcal{U}((g_1, \ldots, g_n )  ,j) \geq \mathcal{U}((g'_1, \ldots, g'_n \},j) 
\] 
for all $g’_j \in G'_j$
\label{def:nash}
\end{definition}

\section{Optimal allocation decision for QC-PRAGM }
\label{sec:optimal-allocation}

To identify the system's optimal performance in QC-PRAGM, we derived the following optimization from equation \eqref{equ:system-cost}: 
\begin{equation}
\begin{array}{ll@{}ll}
\text{maximize}  & \displaystyle\mathcal{S}(\mathcal{G})& &\\
\text{subject to}&&&\\
\text{minimize} \displaystyle\sum\limits_{k = 1}^{q}   &\mathcal{X}_{q_k}f(q_k,  \mathcal{X}_{q_k})&  \\
\text{maximize} \displaystyle\sum\limits_{j = 1}^{n}   &\mathcal{T}(\mathcal{G},j))&  \\
                  &x_{j,k} \geq 0, &j=1 ,\dots, n, k=1 ,\dots, q&\\
                  &z_{j,k} \in \{0, 1\}, &j=1 ,\dots, n, k=1 ,\dots, q&\\
                   &x_{j,k} \leq H.z_{j,k} &j=1 ,\dots, n, k=1 ,\dots, q&\\
                   &\sum_{k=1}^{q} x_{j,k} = r_j & j=1 ,\dots, n& \\
    &\sum_{j=1}^{n} x_{j,k} \leq m_k & k=1 ,\dots, q& \\
\end{array}
\label{eq:equations-with-constraints}
\end{equation}
Indeed,

\begin{equation}
\max_{\mathcal{G}} \; \mathcal{S}(\mathcal{G})
=
\max_{\mathcal{G}}
\left(
-\alpha \sum_{j=1}^{n}\mathcal{F}(\mathcal{G},j)
+
(1-\alpha)\sum_{j=1}^{n}\mathcal{T}(\mathcal{G},j)
\right)
\label{equ:system-cost-opt0}
\end{equation}
\noindent \text{s.t. } $C_1,\; C_2,\; C_3$.

\medskip
\noindent
\textbf{Interpretation (multi-objective).}
Equation \eqref{equ:system-cost-opt0} represents a weighted-sum scalarization of the following bi-objective problem:
\begin{subequations}
\begin{align}
\min_{\mathcal{G}} \; \sum_{j=1}^{n}\mathcal{F}(\mathcal{G},j) \label{eq:sys-a} \\
\max_{\mathcal{G}} \; \sum_{j=1}^{n}\mathcal{T}(\mathcal{G},j) \label{eq:sys-b1}
\end{align}
\label{equ:system-cost-opt1}
\end{subequations}
\noindent \text{s.t. } $C_1,\; C_2,\; C_3$.
Generally, the weighted-sum formulation does not equate to optimizing each objective independently; instead, it chooses a Pareto-optimal trade-off when the feasible set is non-empty. The parameter $\alpha\in[0,1]$ controls the emphasis between cost minimization and performance maximization.

 Finding the resource allocation that minimizes the total client cost in the QC-PRAGM \eqref{eq:sys-a}. This is equivalent to solving a non-linear problem. In fact, \[ \min_{\mathcal{G}} (\sum^n_{j = 1}  \mathcal{F}(\mathcal{G},j)) = \min \sum^{q}_{k=1}\sum^n_{j = 1} \mathcal{E}(j, q_k) f(q_k, \sum^{n}_{j=1}g_j(q_k))\] \[= \min_{\mathcal{G}} \sum^{q}_{k=1} \sum^n_{j = 1} g_j(q_k)f(q_k, \sum^{n}_{j=1}g_j(q_k)) \] \[ \text{subject to} \] \begin{equation}
        \min \sum_{k = 1}^{q} \mathcal{X}_{q_k}f(q_k,  \mathcal{X}_{q_k}), \label{eq:convex}  \end{equation}
         where users send task units to quantum machine $q_k$ with a probability $\mathcal{E}(i, q_k) \in \{0,1\}$. Therefore, the average QPU consumption of quantum node $q_k$ is the sum of its contributions: $\sum^n_{j=1} g_j(q_k) = \sum^n_{j=1} \mathcal{E}(j, q_k)$, and $\mathcal{X}_{q_k} = \sum^{n}_{j=1}g_j(q_k)$.
    This has an optimal solution based on Proof \ref{eq:find-nash-lemma-2}.

 An example output for this problem will be represented as a matrix $M$ that displays numerical values that indicate the number of qubits allocated to each quantum node.  The next step will be to determine which qubit indices need to be placed together to reduce the number of remote gates while respecting the numerical value given in $M$.

  Obtaining the Cost- and inter-node communication-optimized circuit partitioning. We convert the circuits into graphs, where the edges represent the gates. Using matrix $M$, we will group the blocks of qubits of size \(m_{jk}\) that contain the most local gates. These blocks will be assigned to quantum node \(q_k\). This process is described with Algorithm \ref{algo:Inter-node-communication} and will yield an optimal allocation decision that satisfies \eqref{eq:equations-with-constraints}. We have also provided a detailed example \ref{example-cirq-part}.

\begin{prof}
We aim to demonstrate that the resource allocation strategy at the Nash equilibrium corresponds to the optimal resource allocation. From Roughgarden and Tardos~\cite{10.1145/506147.506153} and Jalaparti and Nguyen~\cite{articleJalaparti}, we can deduce the following results:

\begin{lemma}
    Since $X_{q_k}f(q_k, X_{q_k})$ is convex, the local optimal in the non-linear problem 
    \[ \sum_{k = 1}^{q} \mathcal{X}_{q_k}f(q_k, \mathcal{X}_{q_k})\]
    coincides with the global optimal.
    \label{eq:find-nash-lemma-2}
\end{lemma}

\begin{lemma}
Considering QC-PRAGM defined by $(\mathcal{Q}, \mathcal{F}, \mathcal{R})$, then:
\begin{itemize}
    \item There is a Nash equilibrium in QC-PRAGM, and it is unique 
    \item If the cost function $f( )$ ($f: \mathbb{R}^{+} \rightarrow Q\times\mathbb{R}^{+}$). $f(q_k, x)$ is linear, then the total cost at the Nash equilibrium is at most $\frac{4}{3}$ the optimal cost.
\end{itemize}
\label{eq:find-nash-lemma-3}
\end{lemma}

Proof of these results can be done by replacing, respectively, path and edge in \cite{10.1145/506147.506153} with client and QPU.

\end{prof}
\begin{algorithm}
\caption{Cost and Inter-node communication optimizer based on  QC-PRAGM (QC-PRAGM++)}.\label{algo:Inter-node-communication}
\begin{algorithmic}[1]
\Require  A set \( Q = \{ q_1, \ldots, q_q \} \) consists of \( q \) quantum nodes, each defined by the number of computing qubits (\( m_k \)) it possesses. Additionally, there is a list of circuits, referred to as cList, which contains \( n \) circuits. Each circuit requires \( r_j \) computing qubits.
\Ensure the Cost and Inter-node communication optimized circuit partitioning matrix $P = (p_j(q_k))_{0\leq j \leq n, 0\leq k \leq q}$. In this context, $p_j(q_k)$ is a vector that represents the list of qubits from circuit $j$ that need to be allocated within quantum node $q_k$.

\Comment{[Step 1]}
\State Initiaze solver using CVXPY (cp) \cite{diamond2016cvxpy}
\State Decision variables
\Statex \hspace{1em} $X \gets$ cp.Variable(n, q) \Comment{$X = (x_{jk})_{0\leq j \leq n, 0\leq k \leq q}$}
\State Binary indicator for sparsity
\Statex \hspace{1em}$Z \gets$ cp.Variable((n, q), boolean=True) \Comment{Matrix as $X$}
\State Constraints: 
\Statex \hspace{1em} $X \leq H \cdot Z$ \Comment{Big-H sparsity constraint (constraint \( C_1 \))}
\Statex \hspace{1em} $\sum_{k=1}^{q} x_{j,k} = r_j, \;\;\forall j \in \{1,\dots,n\}$ \Comment{constraint \( C_2 \):}
\Statex \hspace{1em} $\sum_{j=1}^{n} x_{j,k} \leq m_k, \;\;\forall k \in \{1,\dots,q\}$ \Comment{constraint $C_3$}
\State Solve: 
\Statex \hspace{1em} \[ \text{objective} \gets \min ( \sum_{k=1}^{q} \sum_{j=1}^{n}x_{j,k}\cdot f(q_k, \sum_{j=1}^{n}x_{j,k}) \]
\Statex \hspace{1em} \text{problem} $\gets$ cp.Problem(objective, constraints)
\Statex \hspace{1em} problem.solver(solver=cp.SCIP)

\State Let $M \gets (m_{jk})_{0 \leq j \leq n, 0 \leq k \leq q}$ 
\State $M \gets X.value$ \Comment{a matrix representing the optimal solution of variable $X$ returned by the solver.}

 \Comment{[Step 2]}
 
\For{$j = 0$ to $n-1$}
\State cgraph $\gets$ Convert cList[j] into graph
    \For{$k = 0$ to $q-1$}
        \State m $\gets$ $M[j,k]$ \Comment{The size of the partition for circuit $j$ to be allocated to quantum node $k$}
        \State sorted\_mutual\_nodes $\gets$ Considering `cgraph`, retrieve the nodes that are the most mutually connected and present them in ascending order
        \State subgraph\_nodes $\gets$ sorted\_mutual\_nodes[:m] 
        \State  $P$[j,k] $\gets$  subgraph\_nodes.qubits
        \State remaining\_nodes $\gets$ nodes - subgraph\_nodes 
        \State cgraph $\leftarrow$ gen\_graph(remaining\_nodes) 
       
\EndFor
\EndFor
\State \textbf{return} $P$\;\\
\end{algorithmic}
\end{algorithm}

\begin{figure}[!t]
    \centering
       

     \includegraphics[width=\columnwidth]{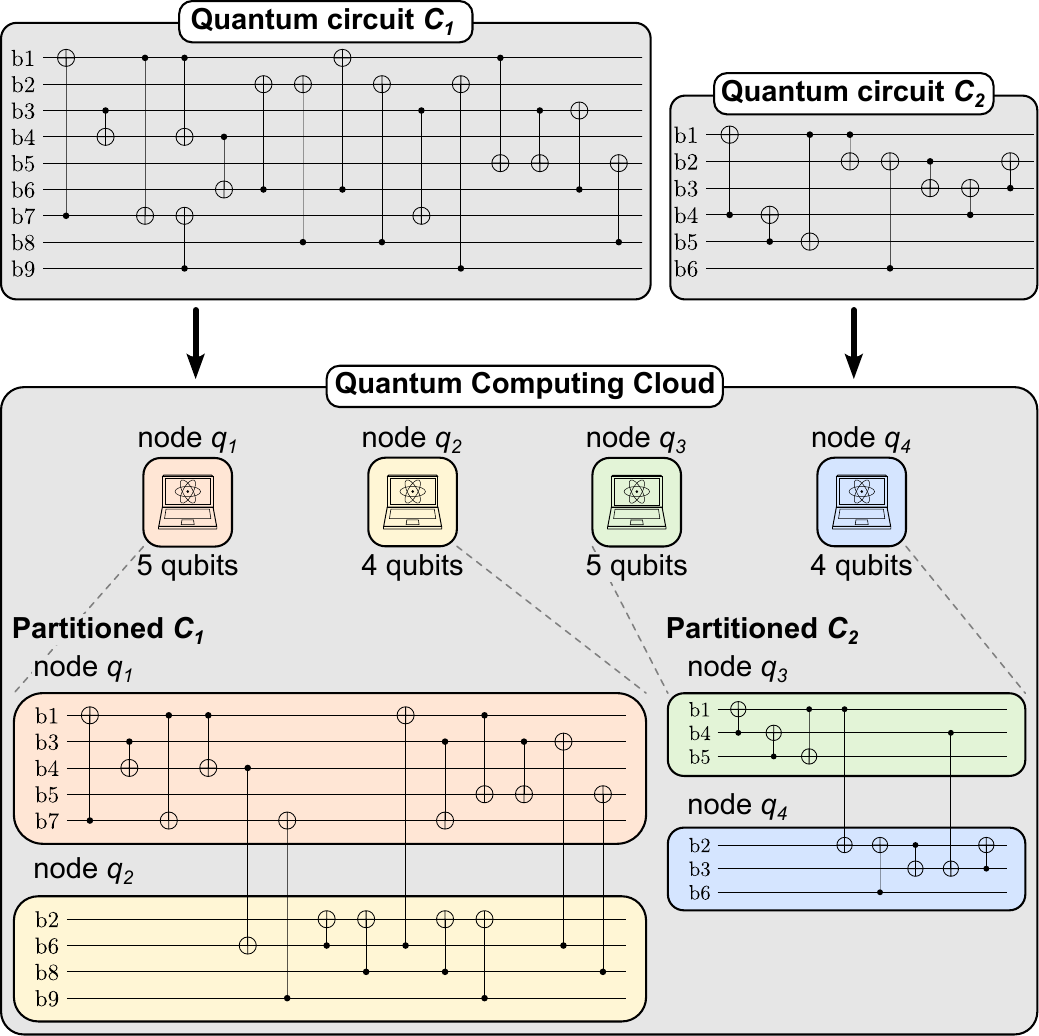}

     
    \caption{\label{fig:example1} We consider four quantum nodes \( Q = \{ q_1, q_2, q_3, q_4 \} \), which have 5, 4, 5, and 4 computing qubits, respectively. Two clients submitted their quantum circuits $\{C_1, C_2\}$, which require $9$ and $6$ qubits to be executed. We define the function \( f: \mathcal{R_{+}} \rightarrow Q\times\mathcal{R_{+}} \) as \( f(q_k, x) = a_{q_k} \cdot x \), which represents the linear cost function associated with quantum nodes. For example, for a node $q_k$ with a capacity of $20$, allocated circuit portions totaling $15$ qubits result in a resource usage of $15/20$ (i.e., $75\%$), which corresponds to the cost $f_{q_k}(75)$. We assume that $a_{q_k} = 100 \$$, which is proportional to the client's fee. After solving our optimization problem ($ \text{minimize} \sum_{k = 1}^{q} \mathcal{X}_{q_k}f(q_k,  \mathcal{X}_{q_k}) = \text{minimize} \sum_{k = 1}^{q} a_{q_k}\sum^{n}_{j=1}g^{2}_j(q_k)$,
    where, $\mathcal{X}_{q_k} = \sum^{n}_{j=1}g_j(q_k)$
) using an open-source Python package named CVXPY \cite{diamond2016cvxpy} (Step 1), we obtained a matrix \( M \) with two rows. The first row, \( (5, 4, 0, 0) \), indicates that for circuit $C_1$, we need to allocate 5 qubits to node $1$ and $4$ qubits to node $2$. The second row, \( (3, 3, 0, 0) \), indicates that for circuit $C_2$, we will allocate $3$ qubits to both node $1$ and node $2$. Step 2 intervenes by transforming $C_1$ and $C_2$ into graphs to identify the combinations of $5$ and $4$ nodes (vertex) in the graph of $C_1$ that have the most mutual edges, referred to as the local gate. The same routine is also exercised on the graph $C_2$. For $C_1$, we obtain the following combinations: $\{b1, b3, b4, b5\}$ with 7 local gates and $\{b2, b6, b8, b9\}$ with 4 local gates. For $C_2$, the combinations are $\{b1, b4, b5\}$ and $ \{b2, b3, b6 \}$, each with 3 local gates.}
    
\end{figure}

\begin{example}
\label{example-cirq-part}  
Example of circuit partitioning based on Algorithm \ref{algo:Inter-node-communication}, where we illustrate two arbitrary quantum algorithms in circuits along with their respective representations after partitioning in Figure \ref{fig:example1}. The resolution of the problem yields a total cost at the Nash equilibrium equal to $27450$ (which corresponds to the optimal cost that the system can achieve), along with the following resource allocation strategy $M$.
\begin{equation}
 \begin{array}{c|cccc}
      &q_1 & q_2 & q_3 & q_4 \\ \hline
   C_1 &5 & 4 & 0 & 0 \\
   C_2 &3 & 3 & 0 & 0
\end{array}
\end{equation}

We then apply Step 2 in Algorithm~\ref{algo:Inter-node-communication}, this process results in the following matrix $P$.
\[
 \begin{array}{c|cccc}
      &q_1 & q_2 & q_3 & q_4 \\ \hline
   C_1 &(b1, b3, b4, b5, b7) & (b2, b6, b8, b9)  & 0 & 0 \\
   C_2 &(b1, b4, b5) & (b2, b3, b6) & 0 & 0
\end{array}
\]

\end{example}

\section{Evaluation}
\label{sec:evaluation}

In this section, we analyze the inter-node communication optimizer using QC-PRAGM++ through simulation. We consider a quantum cloud consisting of 20 quantum nodes, which are fully connected with capacities randomly chosen from the range $(9, 19)$. We compare the performance at the Nash equilibrium with two resource allocation methods: the Round-Robin algorithm and the Random algorithm. Round-Robin involves allocating a quantum circuit to each quantum node (indexed from $0$ to $19$) in a circular order (i.e., a quantum that has previously hosted a quantum circuit will be used again if all quantum nodes have been used). In the Random algorithm, a quantum circuit is assigned to a quantum node chosen randomly. The reason is that these algorithms represent two distinct classes: deterministic and probabilistic, serving as the foundation for various other algorithms. Therefore, we examine various metrics, including the cost per quantum node, the maximum cost, the total system cost, the normalized number of partitions (average partitions divided by the number of circuits, $m$), the normalized number of remote gates (average number of remote gates divided by the number of circuits, $m$), and the latency-induced errors (a function \( L(T_{eg}, T_{cl}, N_{rg}) \), which depends on the latency for entanglement generation, the latency for classical communication, and the number of remote gates) \cite{bahrani2024resourcemanagementcircuitscheduling}.

The Algorithm~\ref{algo:Inter-node-communication} outlines an efficient method for partitioning circuits, focusing on reducing costs and optimizing communication between nodes. We utilize the Kernighan-Lin algorithm \cite{6771089} from NetworkX \cite{SciPyProceedings_11} to perform partitioning for the random and round-robin algorithms. The Kernighan-Lin algorithm can be applied multiple times, depending on the number of partitions needed. In this way, we can find the number of remote gates corresponding to the number of shared edges (gates) between the partition graphs. 

\begin{figure*}[!t]
   \begin{center}

       \subfigure[$m = 10$, Round Robin]{\label{fig:fullyconnected_RR_10_RR_Cost_Per_QPUs}
       \includegraphics[width=0.3\linewidth]{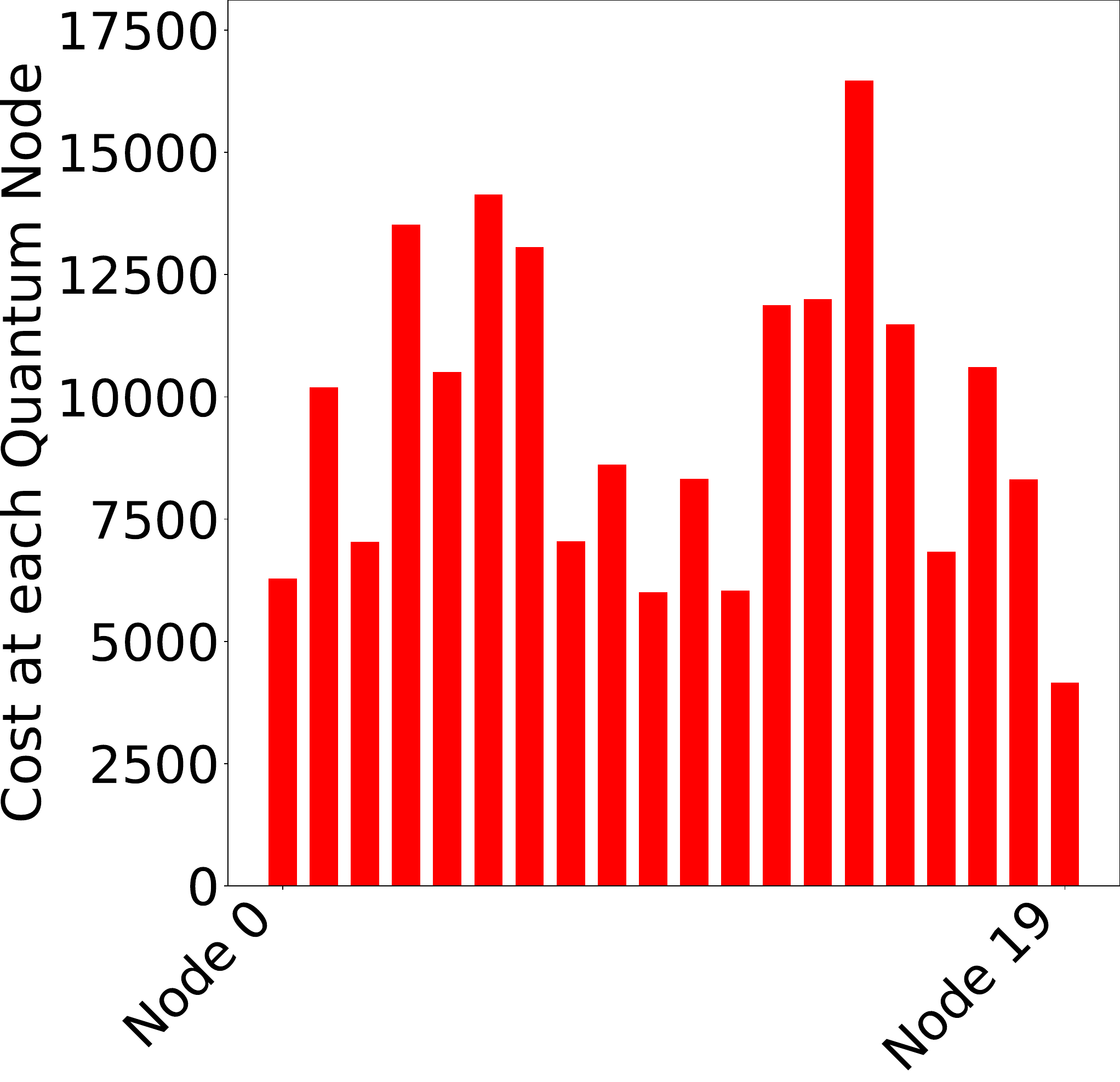}}
       \hspace{0.5mm}\subfigure[$m = 10$, Random]{\label{fig:fullyconnected_Random_10_Random_Cost_Per_QPUs}
       \includegraphics[width=0.3\linewidth]{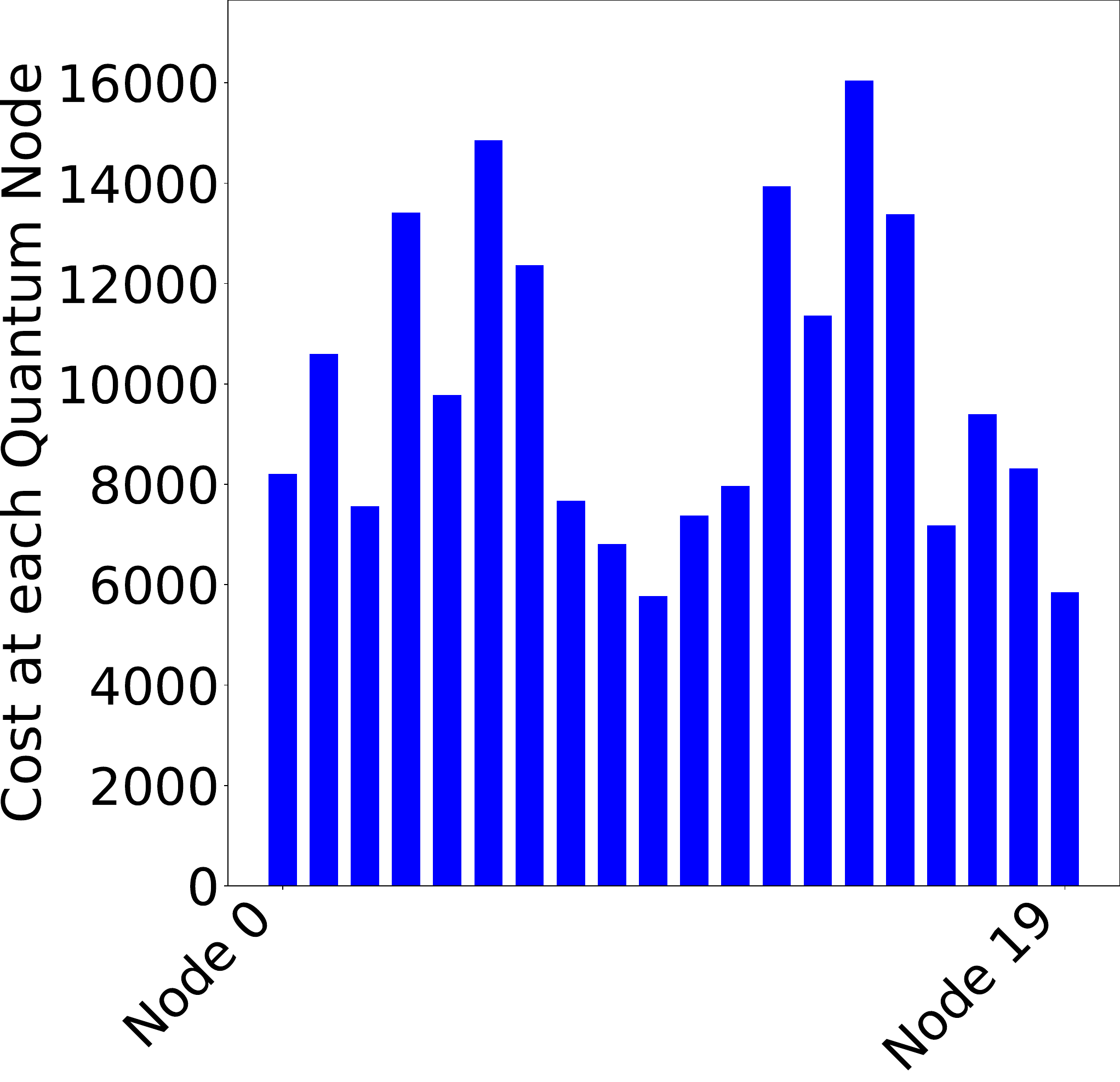}}
       \hspace{0.5mm}
       \subfigure[$m = 10$, QC-PRAGM++]{\label{fig:fullyconnected_PP_10_PP_Cost_Per_QPUs}
       \includegraphics[width=0.3\linewidth]{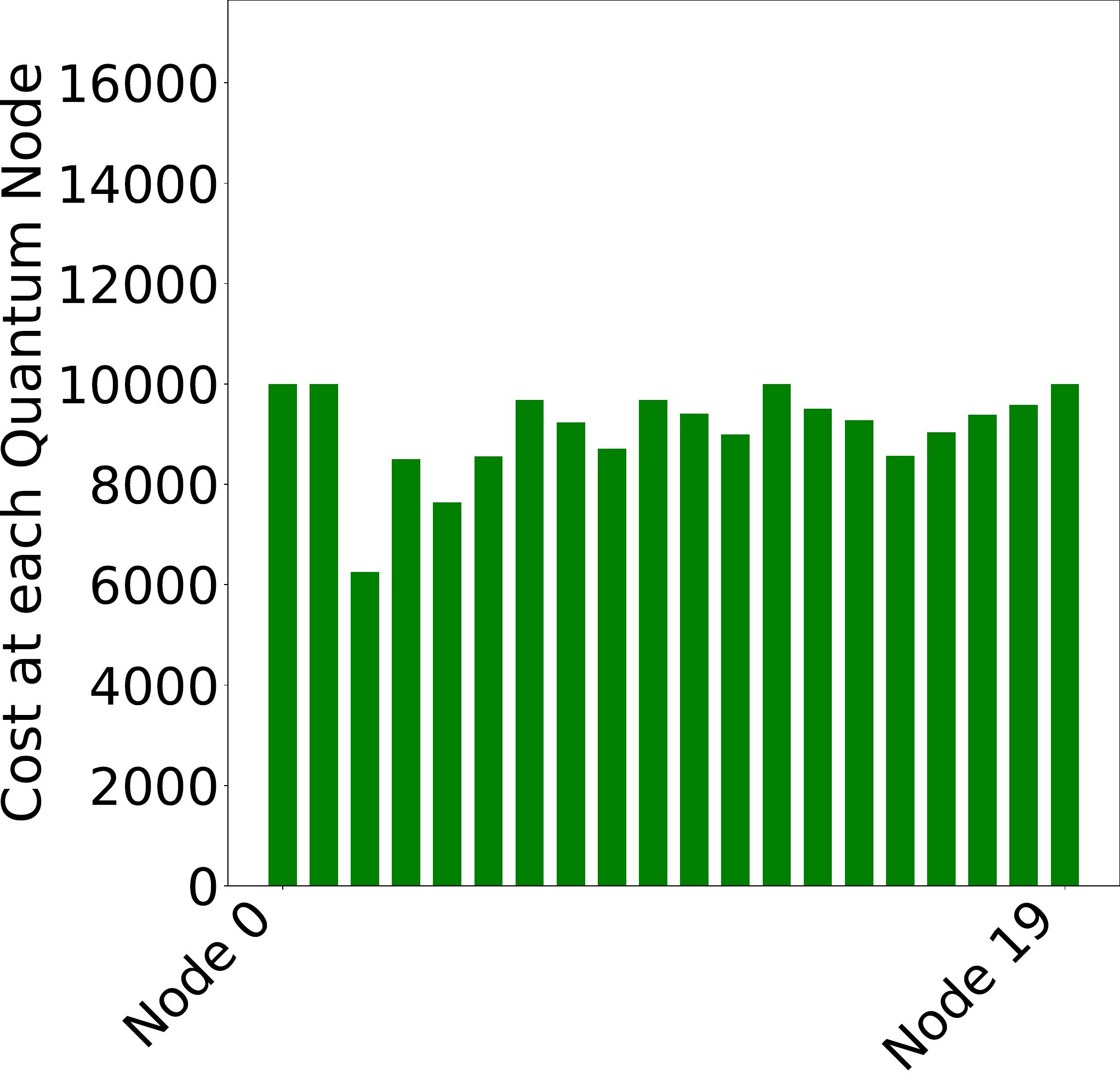}}


   \end{center}

   \begin{center}
   \subfigure[Comparison of System cost ]{\label{fig:ullyconnected_RR_Random_PP_system_cost}
       \includegraphics[width=0.32\linewidth]{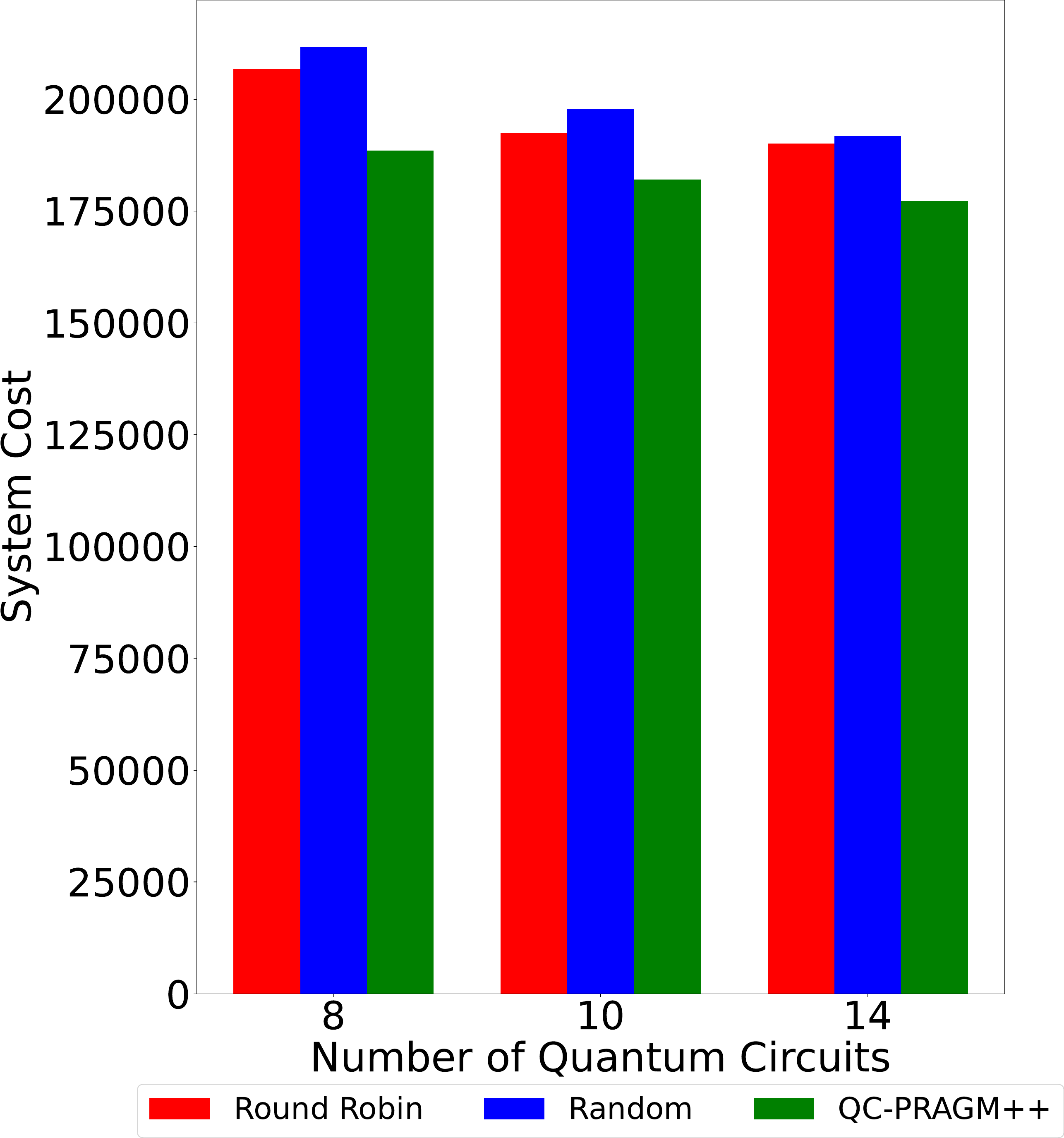}}\hspace{0.0000005mm}
       \subfigure[\text{Comparison of Maximum Costs}]{\label{fig:fullyconnected_RR_Random_PP_max_cost_qpus}
       \includegraphics[width=0.32\linewidth]{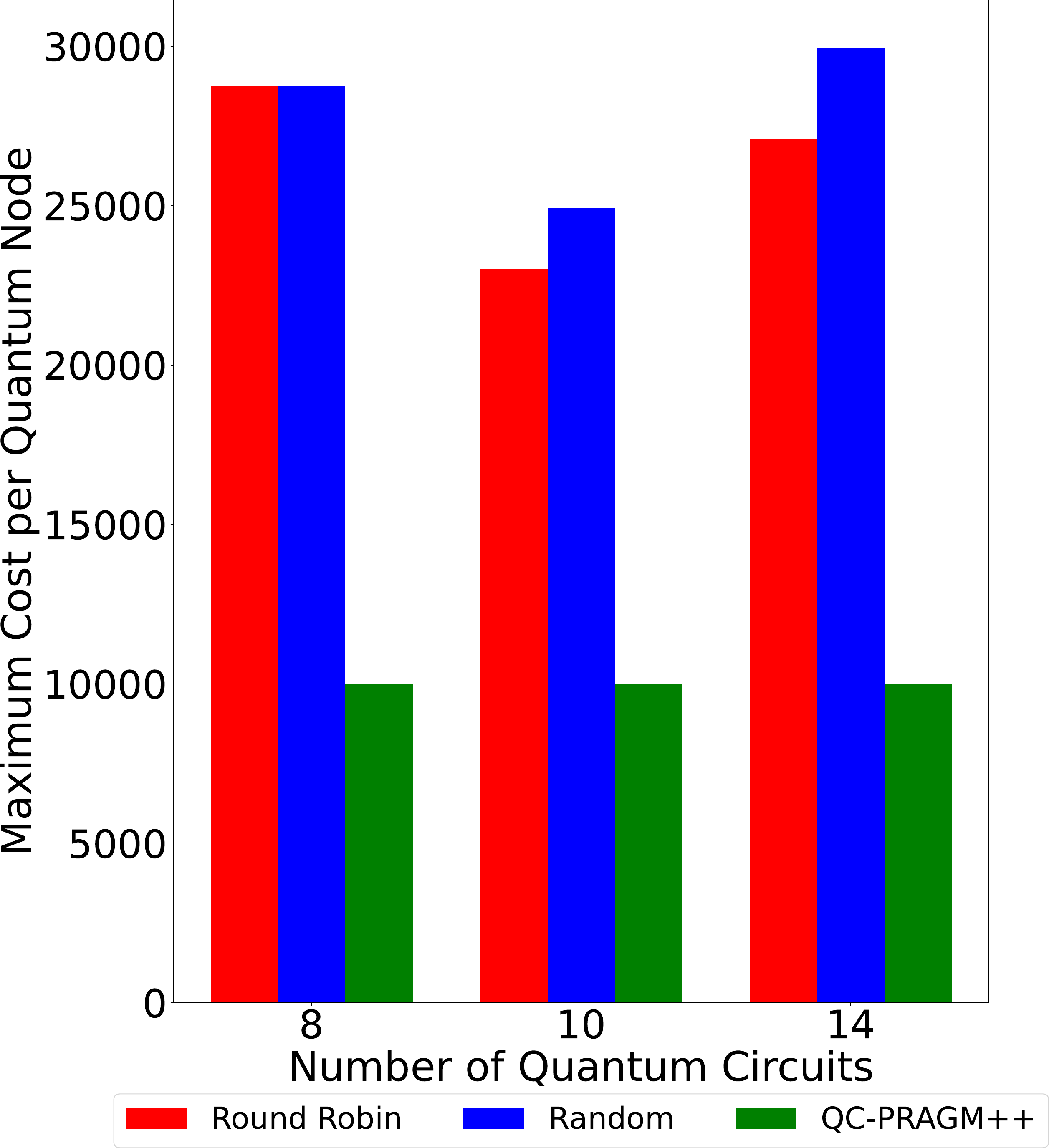}}
       \hspace{0.0000005mm}
       \subfigure[\text{Comparison of partitions}]{\label{fig:fullyconnected_RR_Random_PP_partitions_count}
       \includegraphics[width=0.32\linewidth]{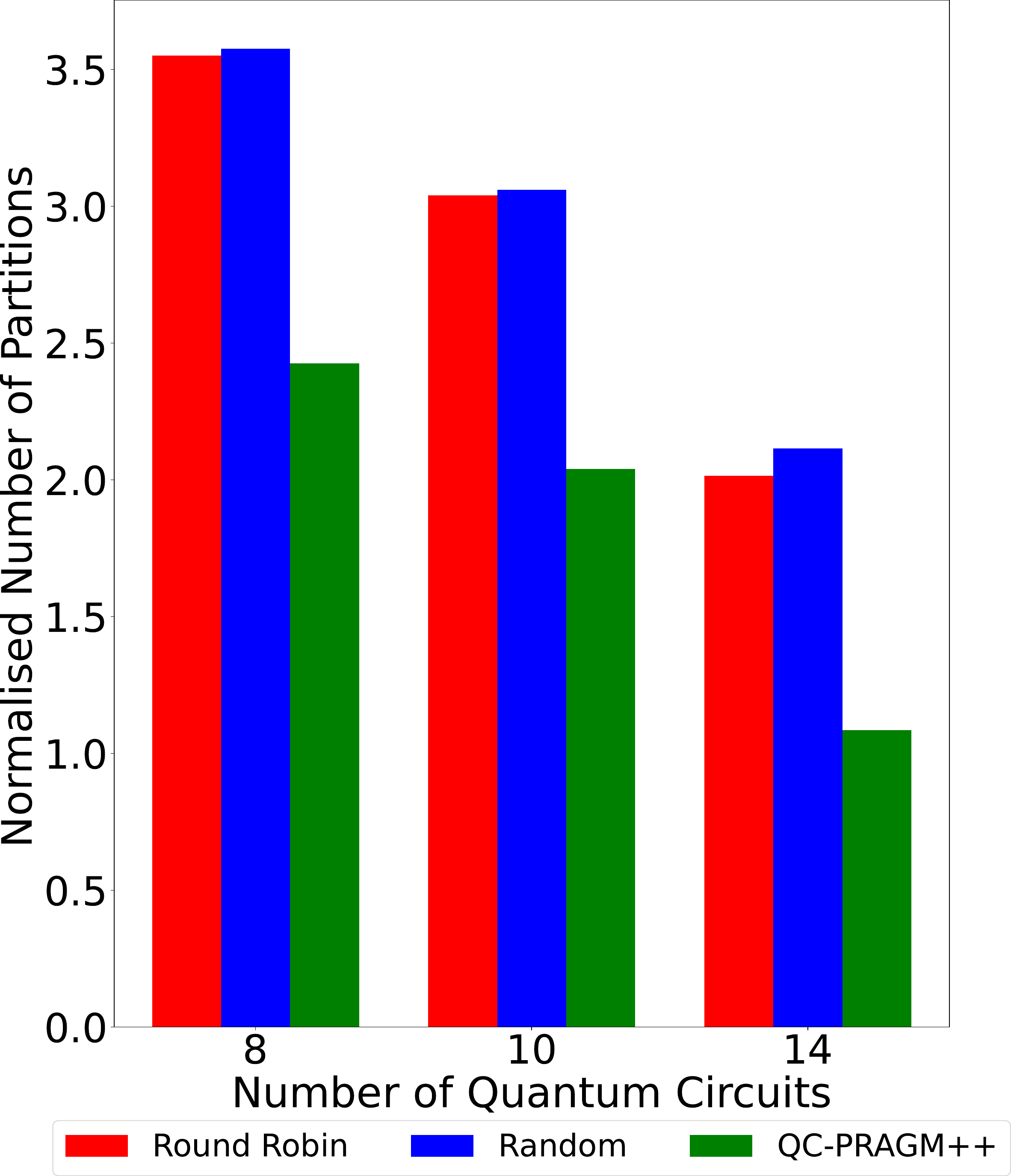}}

   \end{center}

    \begin{center}
       \subfigure[\text{Comparison of remote gates}]{\label{fig:fullyconnected_RR_Random_PP_remote_gates_count}
       \includegraphics[width=0.3\linewidth]{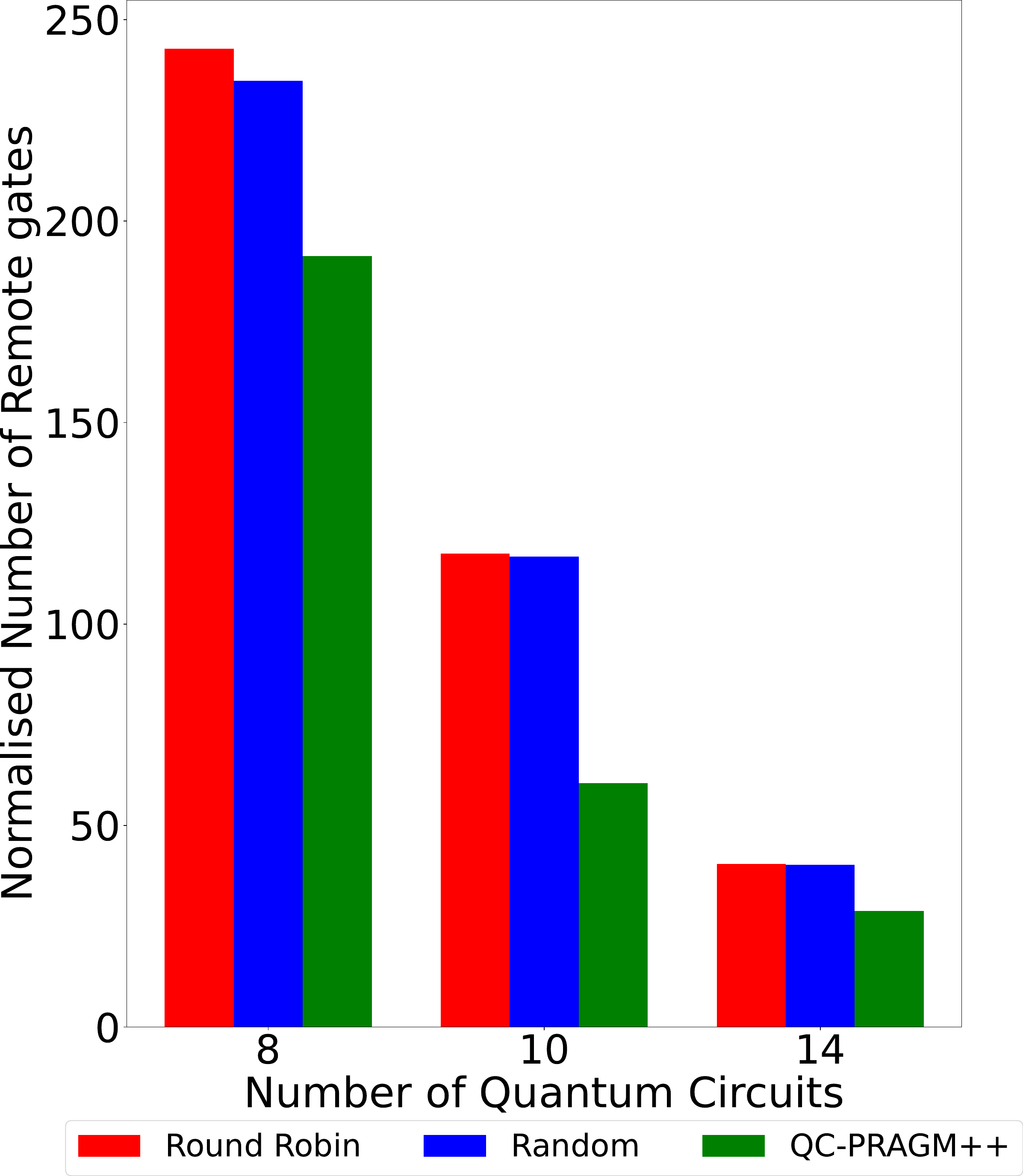}}
        \hspace{0.2mm}
     \subfigure[\text{Comparison of latency-related errors}]{\label{fig:fullyconnected_RR_Random_PP_Latency-Induced Errors-v44}
       \includegraphics[width=0.3\linewidth]{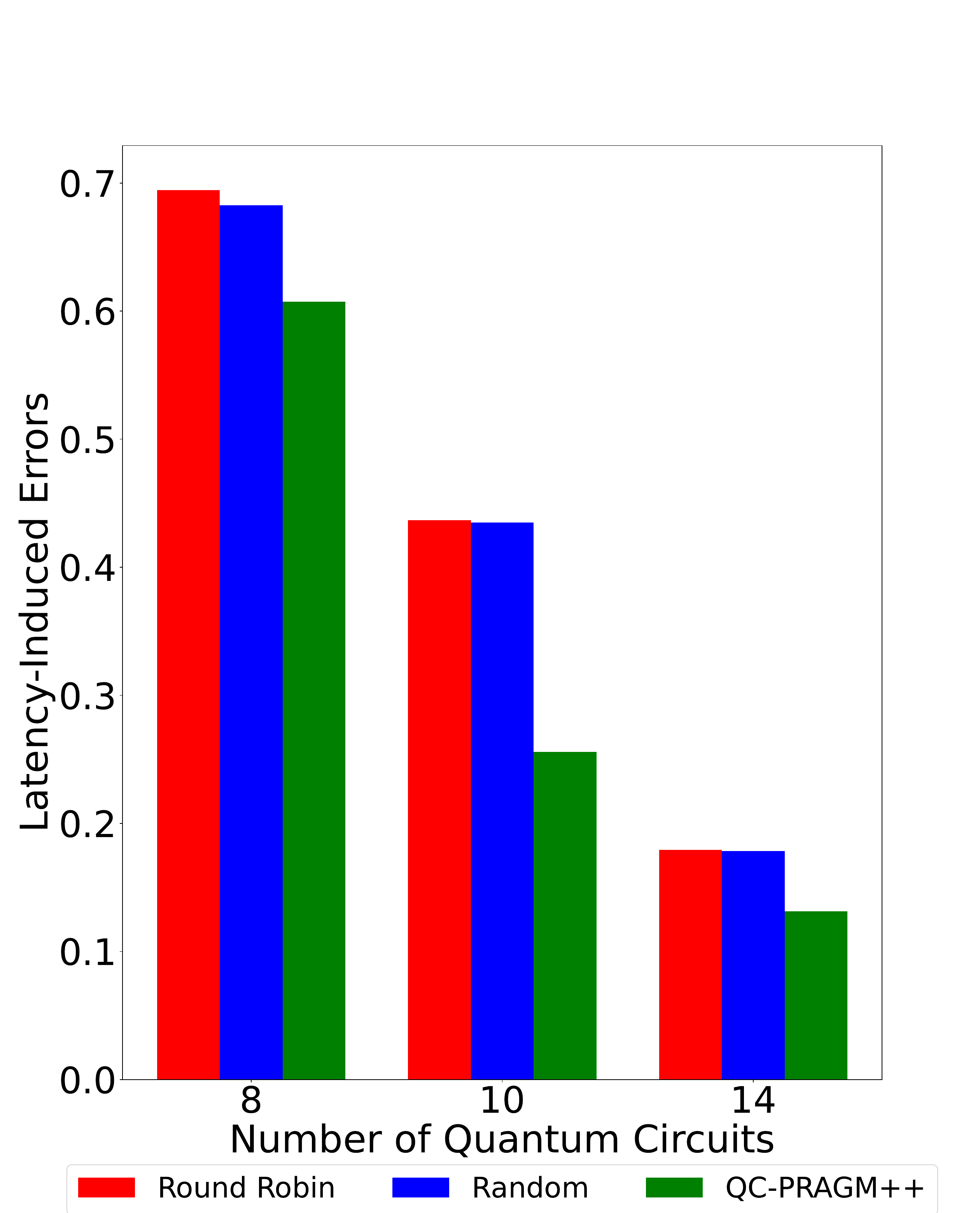}}
   \end{center}

    \caption{Comparison of the cost at each quantum node, system costs, the maximum cost, the number of partitions, the number of remote gates, and the latency-induced error under round-robin, random, and QC-PRAGM++, according to the number of circuits \( m \). The \( m \) quantum circuits are randomly selected from a pool of three distinct types: QFT, DJ, and GHZ state. The set of our quantum circuit is based on the benchmark that the Munich Quantum Toolkit (MQT) \cite{mqt} proposed and is generated using Networkx \cite{SciPyProceedings_11}. As in \cite{bahrani2024resourcemanagementcircuitscheduling}, the widths of these circuits are chosen randomly from the following ranges: for \( m = 8 \), the width is selected from the interval \( (30, 40) \); for \( m = 10 \), from \( (23, 33) \); and for \( m = 14 \), from \( (15, 25) \) and assuming a fully connected network of 20 nodes. Due to limited space, we present the case where \( m = 10 \). }

    
   \label{fig:evaluation}
\end{figure*}


 
\subsection{Performance evaluation}


Figures \ref{fig:fullyconnected_RR_10_RR_Cost_Per_QPUs}, \ref{fig:fullyconnected_Random_10_Random_Cost_Per_QPUs}, and \ref{fig:fullyconnected_PP_10_PP_Cost_Per_QPUs} demonstrate that the QC-PRAGM++ effectively minimizes costs at each quantum node compared to both the round-robin approach and the random method. Indeed, we can observe that, unlike the round-robin approach and the random method, the cost at each quantum node is generally close to the average in QC-PRAGM++. The reason is that QC-PRAGM++ aims to minimize the total cost, resulting in lower expenses for the quantum nodes. This contrasts the round-robin approach and the random method, which allocate as many resources (qubits) as possible to each quantum node, provided its capacity allows it. This can lead to higher costs for some quantum nodes, particularly those with larger capacities. 

In Fig.~\ref{fig:ullyconnected_RR_Random_PP_system_cost}, QC-PRAGM++ demonstrates superior performance compared to both random and round-robin methods regarding system cost. Indeed, the total system cost is minimized for values of $m=8$, $10$, and $14$ when compared to random and round-robin approaches, as shown in Fig.~\ref{fig:ullyconnected_RR_Random_PP_system_cost}. In fact, in QC-PRAGM++, the total cost is reduced by $12\%$, $8\%$, and $8\%$ for \( m = 8 \), \( 10 \), and \( 14 \), respectively, when compared to the random algorithm. Additionally, QC-PRAGM++ results in a total cost reduction of $9\%$, $5\%$, and $7\%$ for \( m = 8 \), \( 10 \), and \( 14 \), respectively, compared to the round-robin algorithm. This confirms the results observed regarding the variation in costs per node in Fig.~\ref{fig:evaluation}. 
Additionally, in Fig.~\ref{fig:fullyconnected_RR_Random_PP_max_cost_qpus}, QC-PRAGM++ shows a maximum uniform cost for $m = 8, 10$, and $14$, which is significantly lower than both the round robin and random methods. Figures \ref{fig:fullyconnected_RR_Random_PP_partitions_count} and \ref{fig:fullyconnected_RR_Random_PP_remote_gates_count} illustrate the variation in the normalized number of partitions and remote gates across all scenarios. QC-PRAGM++ improves communication between the nodes, evident in the reduction of both the normalized number of partitions and the normalized number of gates for $m = 8$, $m = 10$, and $m = 14$. In QC-PRAGM++, circuits are on average partitioned only twice, unlike approaches based on random and round robin, which have twice as many partitions. This result is achieved because our optimizer looks for vectors with a high number of zeros, i.e., while minimizing the cost, we reduce the number of partitions. QC-PRAGM++ also groups together the qubits with the most local gates. Moreover, QC-PRAGM++ performs better than random and round-robin methods in handling latency-related errors, as illustrated in Fig.~\ref{fig:fullyconnected_RR_Random_PP_Latency-Induced Errors-v44}.

\section{Discussions and Limitations}
\label{sec:Discussions-and-Limitations}
Distributed computing is essential to compete with large-scale computing tasks in quantum cloud environments.  The near-unlimited scalability of multicomputer systems and the size of future data centers make it increasingly vital to utilize these resources efficiently. Consequently, effective resource allocation mechanisms are essential in quantum cloud environments. This paper proposes a quantum circuit partitioning resource allocation game model (QC-PRAGM) as an exploratory approach to game theory, a powerful and proven mathematical tool in classical cloud computing, for preventing overcharging customers and minimizing the communication between nodes. We demonstrate analytically that clients are charged appropriately (with a total cost at most $\frac{4}{3}$ the optimal cost) while optimizing quantum cloud resources. Further, our simulations indicate that QC-PRAGM++  performs better than random and round robin approaches in terms of the cost per quantum node, total cost, maximum cost, number of partitions, and number of remote gates. However, we have considered using a fully connected network, which has limited scalability. Therefore, we are considering Q-Fly \cite{sakuma2024opticalinterconnectmodularquantum} topology for the future.

\section*{ACKNOWLEDGMENTS}
Grammarly and Wordtune were used to improve the quality of the text.

\bibliographystyle{IEEEtran}
\bibliography{IEEEabrv,main}
\end{document}